\documentclass[a4paper]{ifacconf}

\usepackage{graphicx,amsmath,url}      
\usepackage[round]{natbib}             

\usepackage{listings}
\usepackage{pgfplots}
\pgfplotsset{compat = 1.11}
\usepackage{pgfplotstable}
\usepgfplotslibrary{groupplots}

\def\portugues{0} 

\ifx\portugues\undefined
\def\portugues{0}
\fi

\if\portugues0
   \usepackage[english]{babel}
  \else
   \usepackage[spanish,brazil,english]{babel}
\fi

\usepackage[T1]{fontenc}

\usepackage[utf8]{inputenc}

\usepackage{ae}

\if\portugues1
 { }
 { }
 { }
 
\fi

\begin{document}

\begin{frontmatter}

\title{Integrating Industrial Artifacts and Agents Through Apache Camel\thanksref{footnoteinfo}} 

\thanks[footnoteinfo]{Sponsor and financial support acknowledgment
goes here. Paper titles should be written in uppercase and lowercase
letters, not all uppercase.}

\author[First]{Cleber Jorge Amaral} 
\author[Second]{Stephen Cranefield} 
\author[Third]{Jomi Fred H{\"u}bner}
\author[Forth]{Mario Lucio Roloff}

\address[First]{Federal Institute of Santa Catarina (IFSC), S{\~a}o Jos{\'e}, Brazil\\(e-mail: cleber.amaral@ifsc.edu.br).}
\address[Second]{University of Otago, Dunedin, New Zealand\\(e-mail: stephen.cranefield@otago.ac.nz)}
\address[Third]{Federal University of Santa Catarina (UFSC), Florian{\'o}polis, Brazil,\\(e-mail: jomi.hubner@ufsc.br)}
\address[Forth]{Instituto Federal Catarinense (IFC), Rio do Sul, Brazil,\\(e-mail: mario.roloff@ifc.edu.br)}

\renewcommand{\abstractname}{{\bf Abstract:~}}   
   
\begin{abstract}                
There are many challenges for building up the \emph{smart factory}, among them to deal with distributed data, high volume of information, and wide diversity of devices and applications. In this sense, Cyber-Physical System (CPS) concept emerges to virtualize and integrate factory resources. Based on studies that use Multi-Agent System as the core of a CPS, in this paper, we show that many resources of the factories can be modelled following the well-known Agents and Artifacts method of integrating agents and their environment. To enhance the interoperability of this system, we use Apache Camel framework, a middleware to define routes allowing the integration with a wide range of endpoints using different protocols. Finally, we present a Camel component for artifacts, designed in this research, illustrating its use. 
\end{abstract}


\end{frontmatter}


\section{Introduction}\label{sec:Introduction}

The ripening of technologies like the Internet of Things (IoT) and Internet of Services (IoS) brought focus of studies on the emergent 4th Industrial Revolution, i.e. Industry 4.0 and Industrial Internet concepts~\citep{CHEN2017588,Kagermann2013,klaus_dieter_thoben_2017_1002731}. These research agree that those technologies are provoking changes on the overall value chain of industries, from raw materials acquisition, logistic, and goods production to the delivery and even post sales services. This wide use and diversity of resources brings challenges when it is necessary to integrate industrial resources as required by the \emph{smart factories} of Industry 4.0 concept. The interoperability, which in this context means that the Cyber-Physical System (CPS) and all sorts of resources can communicate with each other, is a key factor~\citep{Hermann2016,Lu2017a}.

However, the incipient convergence of technology with no standard, usually requires the study of specific integration interfaces and high efforts on development. To face this challenge, we propose the use of Apache Camel framework, a message routing and mediation engine. Camel allows defining communication routes among data producers and consumers by using a Domain-Specific Language, independent of the protocol or networking technology on each side~\citep{Ibsen:2010:CA:1965487}. Camel already has a wide range of components and allows to design new ones facilitating the integration of current and new technologies used by industrial resources.

In many studies~\citep{Leitao2018,Roloff2016,COOK2009277}, Multi-Agent System (MAS) is being the CPS. In fact, it can partially reach some of \emph{smart factory}'s requirements~\citep{Hermann2016,LI2017608,MONOSTORI2016621,Zhong2017}, like decentralization and virtualization, leaving to devices the solution of issues like robustness and real-time capability. However, MAS applications are becoming very complex and computationally heavy specially due indiscriminately \emph{agentification} of entities, i.e., the approach that model almost any entity of a system as an agent. In the other hand, in recent MAS research~\citep{Hubner2010,Omicini2008,Ricci2006,RoloffSHSPH14}, dynamic and complex scenarios are being analysed in dimensions, it interprets that some elements are not necessarily agents. Agents and Artifacts approach (A\&A) is proposing: (i) agent's dimension for proactive entities which encapsulate autonomous execution in some activities (ii) environment dimension which includes \emph{artifacts}, i.e., simpler entities that can be manipulated and shared by agents.

In this sense, the question this paper is employed to answer is: how to integrate all sorts of resources like machines, sensors and software with a MAS in a scalable manner? This paper presents a new component developed to allow the integration of artifacts from MAS to many communication protocols thanks to Apache Camel framework. We think that cognitive agents can play the intelligent part of the system. They can be enhanced with several Artificial Intelligence technologies. The use of artifacts besides a proper method to model non autonomous entities can make the whole system computationally lighter.

This paper is structured as follows: in order to show the background technologies we have used to build a communication component, the framework Apache Camel is briefly presented in Section~\ref{sec:Camel}, and \textsf{CArtAgO} framework, employed to design artifacts, is introduced in Section~\ref{sec:CArtAgO}. Then, in Section~\ref{sec:IndustrialArtifacts}, we show the evolution of factory automation culminating in our proposal and what we refer as \emph{Industrial Artifacts}. In Section~\ref{sec:Component}, we show how the \emph{CamelArtifact} component was modelled and how to use it in an application. Following this, in Section~\ref{sec:Results}, we discuss two illustrative experiments. Finally, related work and conclusions complete this paper.

\section{Apache Camel}\label{sec:Camel}

The Apache Camel framework is a lightweight Java-based message routing and mediation engine~\citep{Ibsen:2010:CA:1965487}. Camel can achieve high performance processes since it handles multiple messages concurrently and provides functions like for interception, routing, exception handling and testing that allows creation of complex routes. This framework uses structured messages and queues as defined on Enterprise Integration Patterns (EIP)~\citep{Hohpe:2003:EIP:940308}, preserving loose coupling among the resources. The complexity of the protocol of each supported technology is embedded in a component, which works as a bridge to Camel routes. There are more than two hundred components available on Camel website and many others on community's repositories. 

A whole route has a data producer, a consumer endpoint, a producer endpoint and, finally, a data consumer. Routes afford the use of multiple endpoints, which means multiple and heterogeneous producers and consumers communicating in the same logical channel where messages go through. An endpoint is a connector that encapsulates a specific protocol. Messages are entities that carry data which has a body, which is the payload, headers and optionally attachments.  \newline

\begin{lstlisting}[caption={Lines 1-3: route from artifact to an external MQTT server to publish in a topic. Lines 5-9: route from a MQTT server to an artifact.}, captionpos=b, label=lst:CamelRoutes, frame=none, framexleftmargin=0pt, numbers=left, numbersep=5pt, numberstyle=\tiny, language=Java, showspaces=false, showtabs=false, breaklines=true, showstringspaces=false, breakatwhitespace=true, basicstyle=\footnotesize\ttfamily,
morekeywords={route, from, to, setHeader}]
from("artifact:cartago")
.transform().mvel("(request.body[0] * 1.8 + 32).toString()")
.to("mqtt:foo?host=tcp://broker(...)");

from("mqtt:foo?host=tcp://broker(...)")
.setHeader("ArtifactName",constant("s1")) 
.setHeader("OperationName",constant("temp"))
.transform().mvel("[ (request.body[0].toString() - 32) / 1.8 ]")
.to("artifact:cartago");
\end{lstlisting}

In order to write route definitions, there are three available Domain-Specific Languages (DSLs): Java, Scala and XML based language. Using these languages, Camel allows to wrap in the route the necessary transformations to integrate a set of data consumers and producers. Camel works as a middleware that can be incorporated in an application for concentrating integration matters. In this fashion, programming complexity may be reduced since there is a separation between MAS and integrating programming.

For instance, a route may be used to convert temperature unities when an endpoint that uses celsius needs to send some data to another endpoint that is expecting it in fahrenheit. The code in Listing~\ref{lst:CamelRoutes} illustrates its route definitions in Java. In the first, the route is processing the content, applying some math and sending to a MQTT\footnote{Message Queuing Telemetry Transport by IBM\texttrademark.} endpoint. The next route is displaying the way back, adding on the header of the message necessary tags to destination endpoint\footnote{broker(...) refers to broker's address. Some Math is omitted.}.


\section{\textsf{CArtAgO} Artifacts}\label{sec:CArtAgO}

In MAS the agents are situated entities. They are perceiving and acting on an environment. Non-agent elements of a MAS are usually considered as part of the environment~\citep{Weyns:2007:EFC:1176841.1176951}, which may have tools and other objects that can be used and shared by agents. The framework \textsf{CArtAgO} calls these resources \emph{artifacts}~\citep{Ricci2006}. Essentially what differs Agents and Artifacts is autonomy, agents are considered the active part of the system. The Artifacts, on the other hand, are not autonomous, they have functions, they provide operations and behave predictably~\citep{Omicini2008}. 


Artifacts commonly are utilized to: (i) simulate the real world, (ii) as virtual representations touchable by the agents, (iii) for coordination purposes, and (iv) as interfaces to the external world wrapping some technology. Many of these uses are related to shareable knowledge about the environment, which refers to synchronization issues. About wrapping functions, the called \emph{resource artifacts} are responsible for this. They mediate access to such functions or effectively embody a resource of a MAS. 

\begin{figure}[ht]
\centerline
{
    \includegraphics[width=0.32\textwidth]{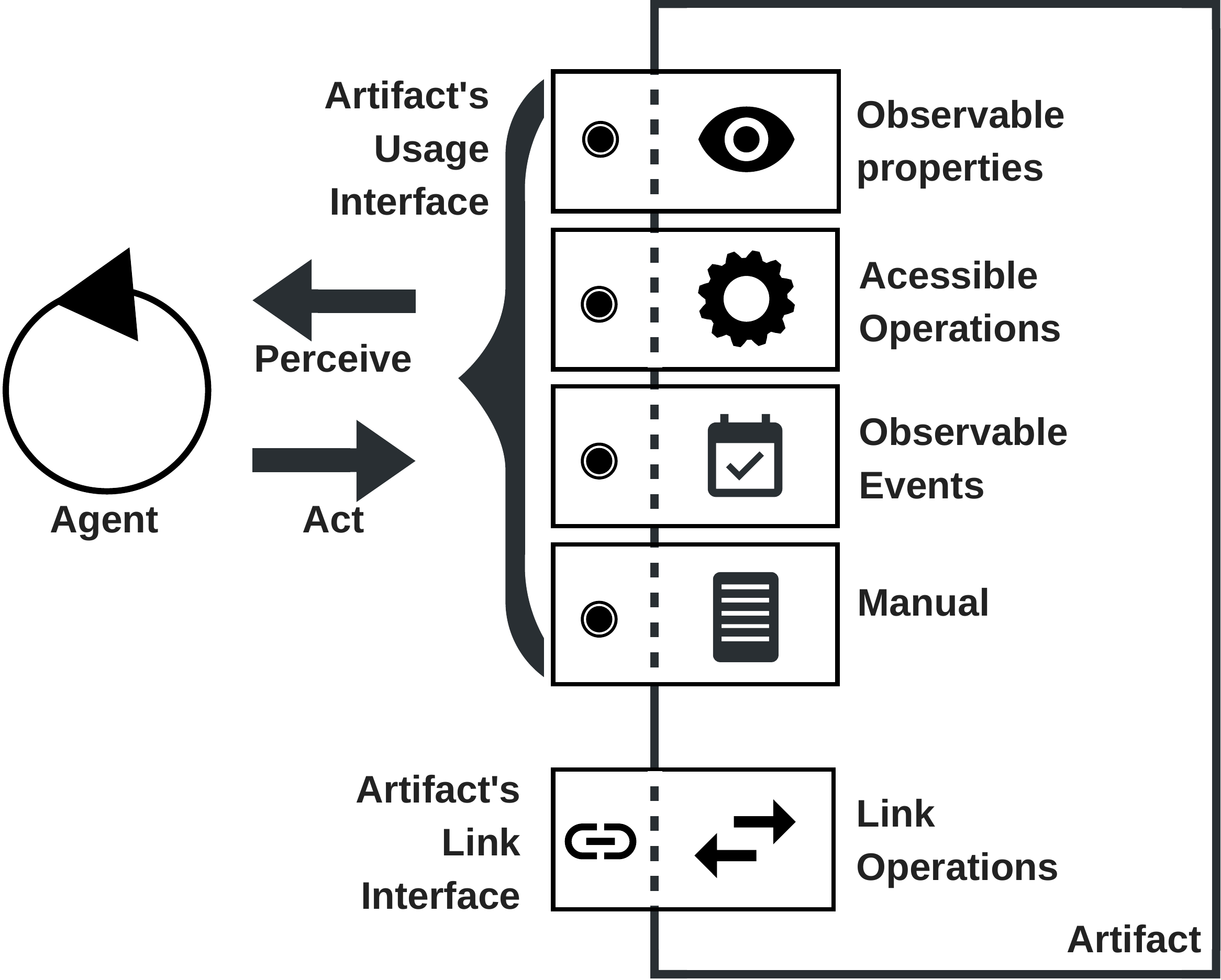}
}
\caption{Artifact's structure.}
\label{fig:AABAsicLevel}
\end{figure}


\textsf{CArtAgO} is a Java-based framework that brings many functions to promote knowledge synchronization among agents and environment. The API provides basic classes to define artifacts, the interface to interact with agents and a run-time infrastructure to support the dynamic management of working environments~\citep{Ricci2006}. Besides these facilities, the framework processes transactions atomically to ensure data integrity and provides synchronization functions for multiple agents and multiple infrastructure servers.

Artifacts are used by the agent through its interface which provides operations to achieve the services it offers. Typically, artifacts' implementation are computationally lighter than agents since they are passive entities. The operations performed by artifacts commonly require little attention from the agent since they are passive and characterized to be routine behaviour. 

Artifacts are located in logical areas called \emph{workspaces}. The agent that is focusing on an artifact reads its \emph{observable properties}, perceives events and may trigger its interfaced operations. Another feature that artifacts may provide is a manual with machine-readable description of their functions which is useful especially in open systems. Finally, artifacts provide linking interfaces allowing to connect artifacts and use linked operations (Figure~\ref{fig:AABAsicLevel}). With this function an artifact may invoke an operation of another, for instance, to communicate with a resource through another artifact.

\section{Industrial Artifacts}\label{sec:IndustrialArtifacts}

\begin{figure}[ht]
\centerline
{
    \includegraphics[width=0.32\textwidth]{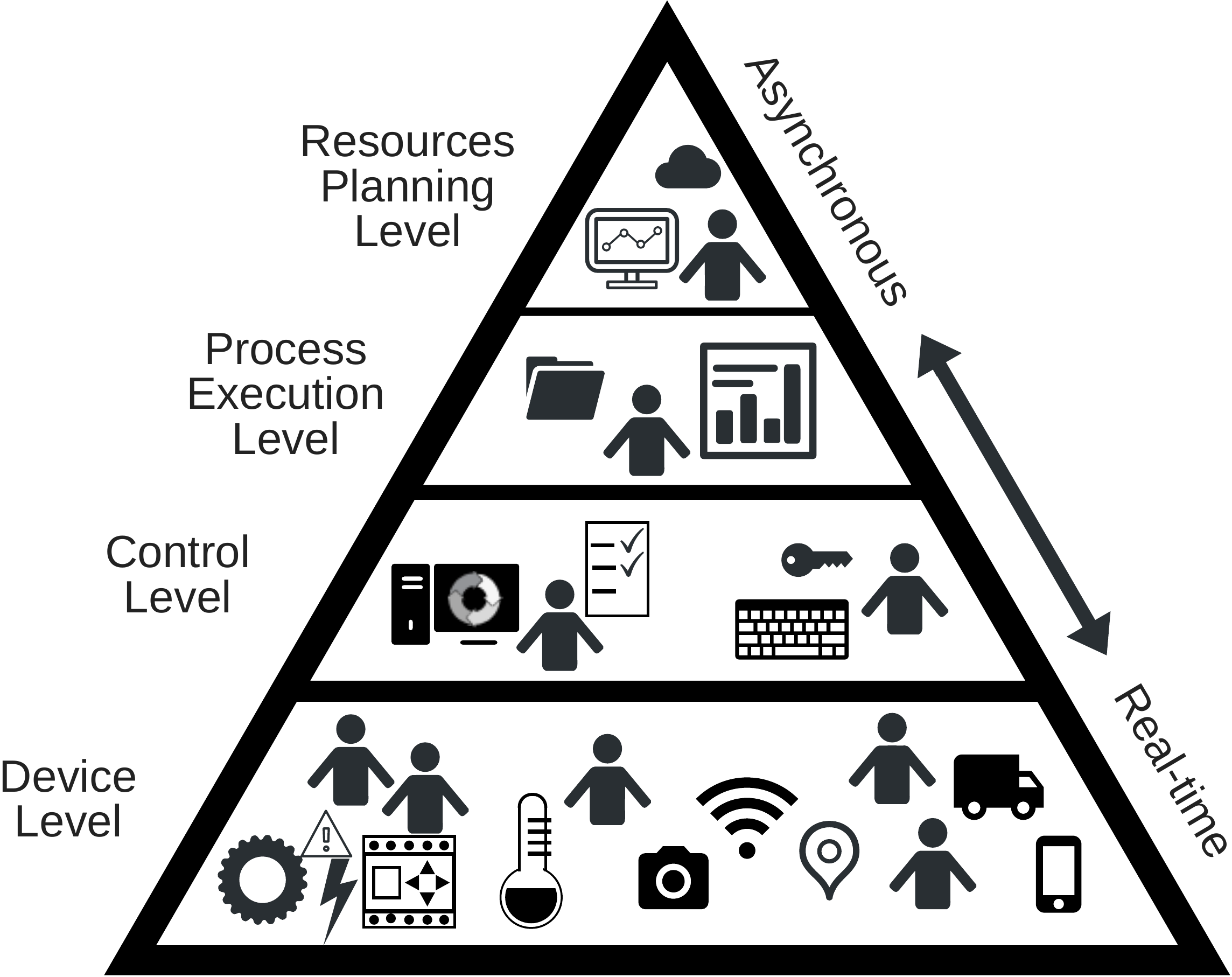}
}
\caption{Traditional Automation Pyramid}
\label{fig:pyramid}
\end{figure}

A traditional automated factory may be seen in four levels of a pyramid as showed in Figure~\ref{fig:pyramid}. The strategic decisions are placed on an Enterprise Resource Planning level. According to priorities and other aspects, the manufacturing is scheduled and monitored in the Process Execution level. The next level has the responsibility to control the wide spread devices, often in real-time. These devices, like sensors and actuators are situated in the bottom level of the pyramid. The presented pyramid is being more populated for all sorts of nodes including peripheral nodes (i.e logistic monitoring and control). The automation is usually still partial as we can see by human presence, in all levels, filling gaps of the processes.

\begin{figure}
\centerline
{
    \includegraphics[width=0.425\textwidth]{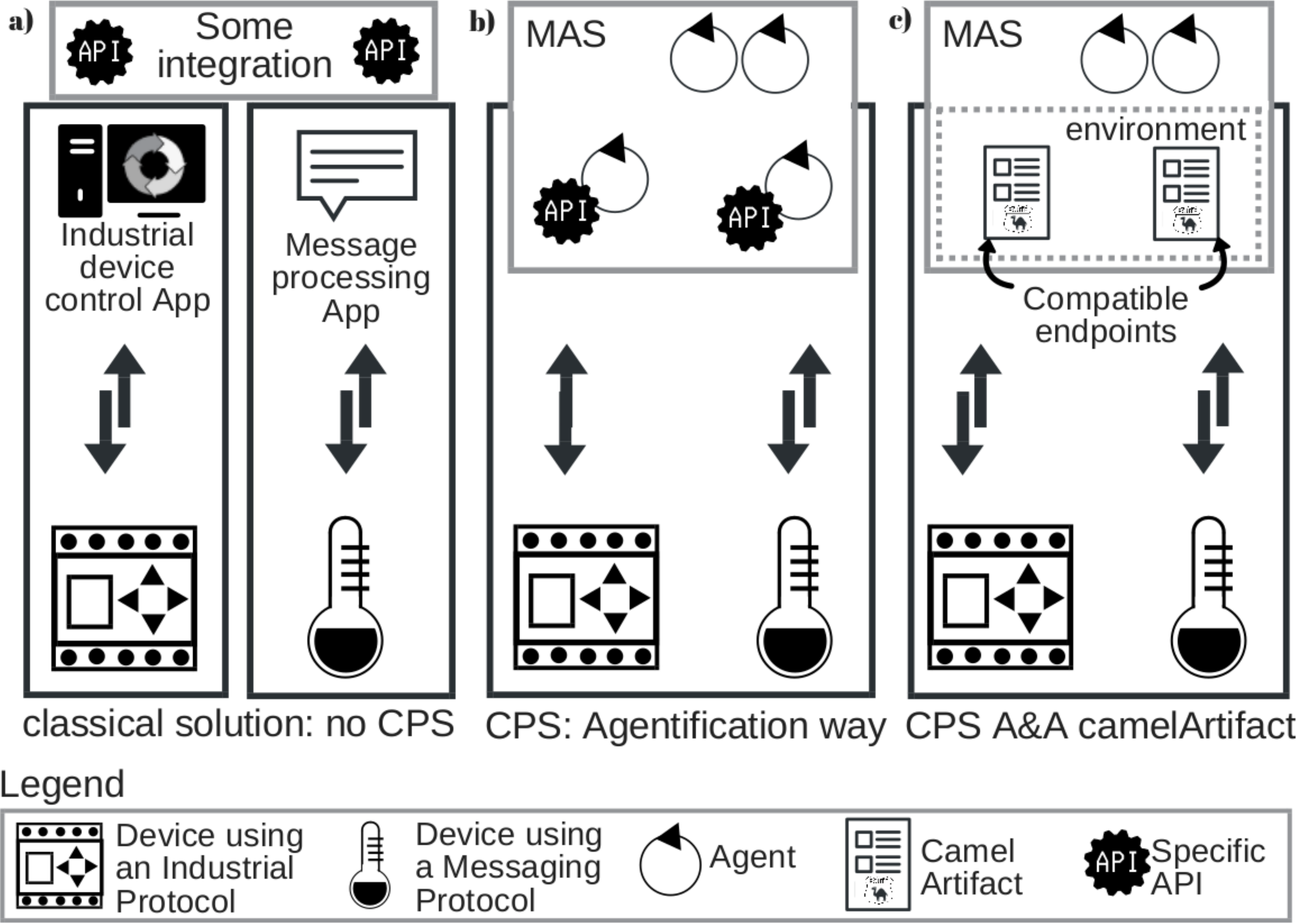}
}
\caption{A factory automation integrating two devices. a) ad hoc solution. b) \emph{agentification} approach. d) CamelArtifact and endpoints.}
\label{fig:Comparison}
\end{figure}

Many of those devices communicate via OPC (OLE for Process Control) which is widely accepted communication standard on factory shop floor automation\citep{Heory2014}. This standard is used by devices and supervisory software located on Device and Control levels of the pyramid. The higher levels of the pyramid commonly use other applications with restricted or no integration with lower levels. In this scenario, as illustrated in Figure~\ref{fig:Comparison}a, interoperability among different levels and different technologies is commonly solved by ad hoc solutions. The figure illustrates some integration between an industrial device (i.e. an OPC controller) and a messaging device (i.e. IoT sensor) by specific APIs (Application Program Interfaces).

Factory automation is evolving, especially towards Cyber-Physical Systems (CPS) concept, which is integrating virtual and physical processes~\citep{4519604,LI2017608}. CPS has a standardized abstraction and architecture integration in a broad sense~\citep{MONOSTORI2016621}. This concept is central to the so called \emph{smart factory} of the Industry 4.0~\citep{Kagermann2013}. MAS is playing the central part of the CPS virtualizing entities allowing decentralized control and interoperability in many researches. However, studies~\citep{Marik:2005:RAA:1082473.1082812,Yokogawa2000MachineryCS} opted to represent almost any factory entity as agents (Figure~\ref{fig:Comparison}b), which increases complexity and makes synchronization of the environment information more difficult.

Virtual representations of the plant, including software entities as well as the physical world, may be reached at A\&A approach using \emph{artifacts} and \emph{workspaces}. All sorts of resources that accept command by operations and generate events can be modelled as artifacts. Artifacts allow to represent heterogeneous entities in a common format and make these virtual representations interoperable. In fact, without a mediation tool like Camel the usual solution to integrate artifacts and each technology is by APIs, what increases development efforts. Using our proposition (Figure~\ref{fig:Comparison}c) interoperability is facilitated using Camel that makes available many components for different protocols. The integration provided by Camel is also facilitated through DLSs, the application just need to specify the routes via URIs parameters and message headers. In most cases this approach provides the needed functionality with fewer programming efforts and faster learning curve.

\section{The Artifact Component}\label{sec:Component}

In order to address interoperability, we propose joining Camel framework and \emph{artifacts}. Artifact's operations are responsive execution processes lighter than agent's actions which are usually consciously behaviour. Different device protocols and networking technologies can be integrated to a MAS similarly, using Camel as routing and mediation engine. In this sense, the developed component, called \emph{CamelArtifact}, is responsible for link artifacts and external resources. Each \emph{CamelArtifact} may contain routes definitions using specific endpoints for each resource. The endpoints are encapsulating the communication protocol complexity. 

To use this component, an instance of \emph{CamelArtifact} should be set to listen its routes. Any message coming from, or going to the routes, will be kept in queues. Messages that are arriving or being sent can be transformed. The transformation is usually to make compatible both sides of the route depending on the application. The artifact may have route definitions on itself or it may receive and send messages through other \emph{CamelArtifact}, which may forward messages. The use of forwarding function may save computer memory, on the other hand, each artifact as a \emph{CamelArtifact} has his own thread taking computer parallelism advantages. 



\subsection{The Component Architecture}

The structure of the created component was based on the available \emph{Camel Component Archetype}, which provides a useful component template. The default configuration of an Apache Camel component is mainly a \emph{DefaultComponent} class that creates its endpoints, normally a consumer and a producer. In a MAS application, an artifact essentially uses \textsf{CArtAgO} Artifact original class. In our component, a new class, the called \emph{CamelArtifact}, extends Artifact from \textsf{CArtAgO} framework and imports the Camel API to implement Apache Camel routes. The data used by the artifacts were modelled as \textit{OpRequest}, meaning Operation Request, which contains the name of an artifact, an operation to be performed and its parameters. 


Producer and consumer sides work in a very similar manner by polling processes supported by queues for incoming and outgoing messages. With the polling consumer to send messages, the artifact places messages in the outgoing queue that will be repeatedly checked by the endpoint consumer to send it through the route. For incoming messages process, the component implements an ad hoc process using CArtAgO's IBlockingCmd class to check for new messages delivering it to the artifact.

The message structure used by Camel provides in the header a map of Java objects, in the body any Java object, and optional attachments. In the header of \emph{CamelArtifact} messages, it is expected to find tags for the artifact name and operation, this last regards to a method to be performed. The body may contain a list of params the referred method needs. The operation tagged will be invoked as a \textsf{CArtAgO} internal operation or it may be forwarded when addressed to other artifact. 

\section{Illustrative use and Results}\label{sec:Results}

To test the designed Camel component, two MAS applications were developed. The first was centred on a scalability experiment of the component making use of multiple consumers, forwarding function and \emph{loopback} communications. The second experiment was to try the component in a scenario needing interoperability among different technologies and protocols in the context of Industry 4.0.

\subsection{Terminal and Router scalability experiment}

In the first application, two scenarios are used to illustrate the \emph{ArtifactComponent}. For communication point of view, there are the scenarios have artifacts as terminals, which are end-points of the communication and an artifact as a router, which is a middleware to forward messages to end-points. Scenario 1 is varying the number of terminals instantiating multiple \emph{CamelArtifacts}. Scenario 2 is varying the number of common artifacts, all of them linked to a \emph{CamelArtifact} as a message router. The router, in this context, contains routes of other artifacts using Apache Camel MQTT supported endpoint. 

All the \emph{CamelArtifacts} were set to publish and subscribe its own topic, being able to receive back its own sending messages. When playing the router function, an extra route was created to the linked artifacts. The scenario 1 had the number of camel terminals varied from 10 to 500 artifacts. The scenario 2 had the number of common artifacts varied from 10 to 200.

The MAS was designed with Jason framework~\citep{Boissier:2013:MOP:2459520.2459792}. It has only one agent, liable to make the artifacts, link and manage them. We have used MQTT QoS 2 setting, which mean the most guaranteed message delivering method provided by this protocol. The resources exchanged messages in both ways every 6 seconds. For our virtual Ubuntu Linux server with 2 cores and 2 GB RAM, it is a stressing situation that allows to check the message processing limits. 

In scenario 1, we notice that the system uses more RAM memory since it creates several Camel instances. The application with 10 \emph{CamelArtifacts} used 127 MB growing an average of 1.3 MB for each instance added. In scenario 2, variance is not conclusive, main changes occurred by other Java Virtual Machine processes~(Figure~\ref{fig:Results}a). In addiction, on scenario 1 the system needed more time to load, for 10 artifacts it needed 21 seconds increasing an average of 1.5 seconds for each \emph{CamelArtifact} added. Scenario 2 had no significant increase~(Figure~\ref{fig:Results}b). In contrast, messages per second rate of scenario 1 reach better results, growing from 2 to 16.6 when the system was tested with 200 \emph{CamelArtifacts}, scenario 2 grew only until 4.1 messages/s~(Figure~\ref{fig:Results}c). 



\begin{center}
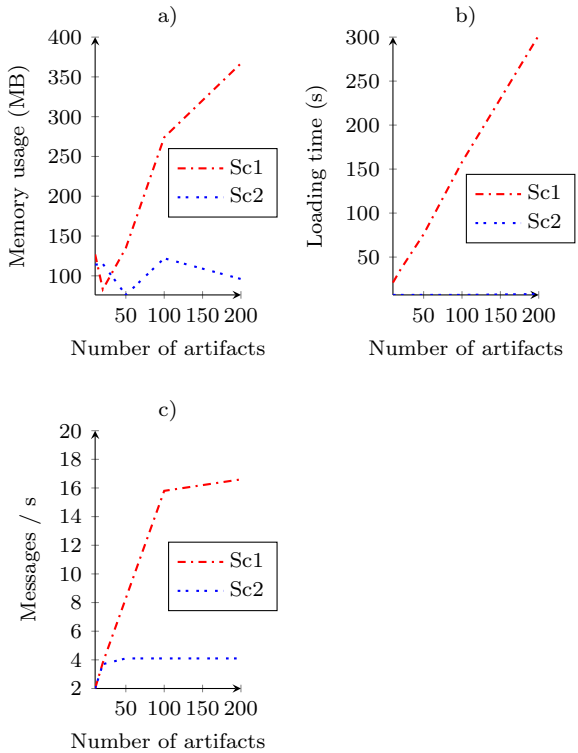
\begin{figure}[h]
\begin{tikzpicture}
\begin{groupplot}[footnotesize,
	group style={group size=2 by 2,
       	horizontal sep=2.0cm,
        vertical sep=1.8cm},
        height=5.0cm,width=3.5cm,
]
\nextgroupplot[
	axis x line=bottom,
    axis y line=left,
    ymax = 400,
    xmax = 200,
    xlabel = Number of artifacts, 
    ylabel = Memory usage (MB),
    title = {a)},
    legend style={at={(0.5,0.3)},anchor=south west}
]
\addplot[thick, red, dash dot] 
	table[y=Scenario1,x=nArtifacts,col sep=comma]{Experiment1Memory.csv};
\addplot[thick, blue, dash pattern={on 1pt off 2pt on 1pt off 3pt}]
	table[y=Scenario2,x=nArtifacts,col sep=comma]{Experiment1Memory.csv};
\addlegendentry{Sc1}
\addlegendentry{Sc2}

\nextgroupplot[
	axis x line=bottom,
    axis y line=left,
    ymax = 300,
    xmax = 200,
    xlabel = Number of artifacts,
    ylabel = Loading time (s),
    title = {b)},
    legend style={at={(0.5,0.2)},anchor=south west}
]
\addplot[thick, red, dash dot] 
	table[y=Scenario1,x=nArtifacts,col sep=comma]{Experiment1LoadingTime.csv};
\addplot[thick, blue, dash pattern={on 1pt off 2pt on 1pt off 3pt}]
	table[y=Scenario2,x=nArtifacts,col sep=comma]{Experiment1LoadingTime.csv};
\addlegendentry{Sc1}
\addlegendentry{Sc2}

\nextgroupplot[
	axis x line=bottom,
    axis y line=left,
    ymax = 20,
    xmax = 200,
    xlabel = Number of artifacts,
    ylabel = Messages / s,
    title = {c)},
    legend style={at={(0.5,0.3)},anchor=south west}
]
\addplot[thick, red, dash dot] 
	table[y=Scenario1,x=nArtifacts,col sep=comma]{Experiment1nMsgsPerSec.csv};
\addplot[thick, blue, dash pattern={on 1pt off 2pt on 1pt off 3pt}]
	table[y=Scenario2,x=nArtifacts,col sep=comma]{Experiment1nMsgsPerSec.csv};
\addlegendentry{Sc1}
\addlegendentry{Sc2}

\end{groupplot}
\end{tikzpicture}
\caption{Scenarios 1 and 2 (Sc1 and Sc2) comparison as increase the number of artifacts. a) Memory use b) Time to load c) Message/s rate }
\label{fig:Results}
\end{figure}
\end{center}

\subsection{Industry 4.0 context experiment}

\begin{figure}[ht]
\centerline
{
    \includegraphics[width=0.32\textwidth]{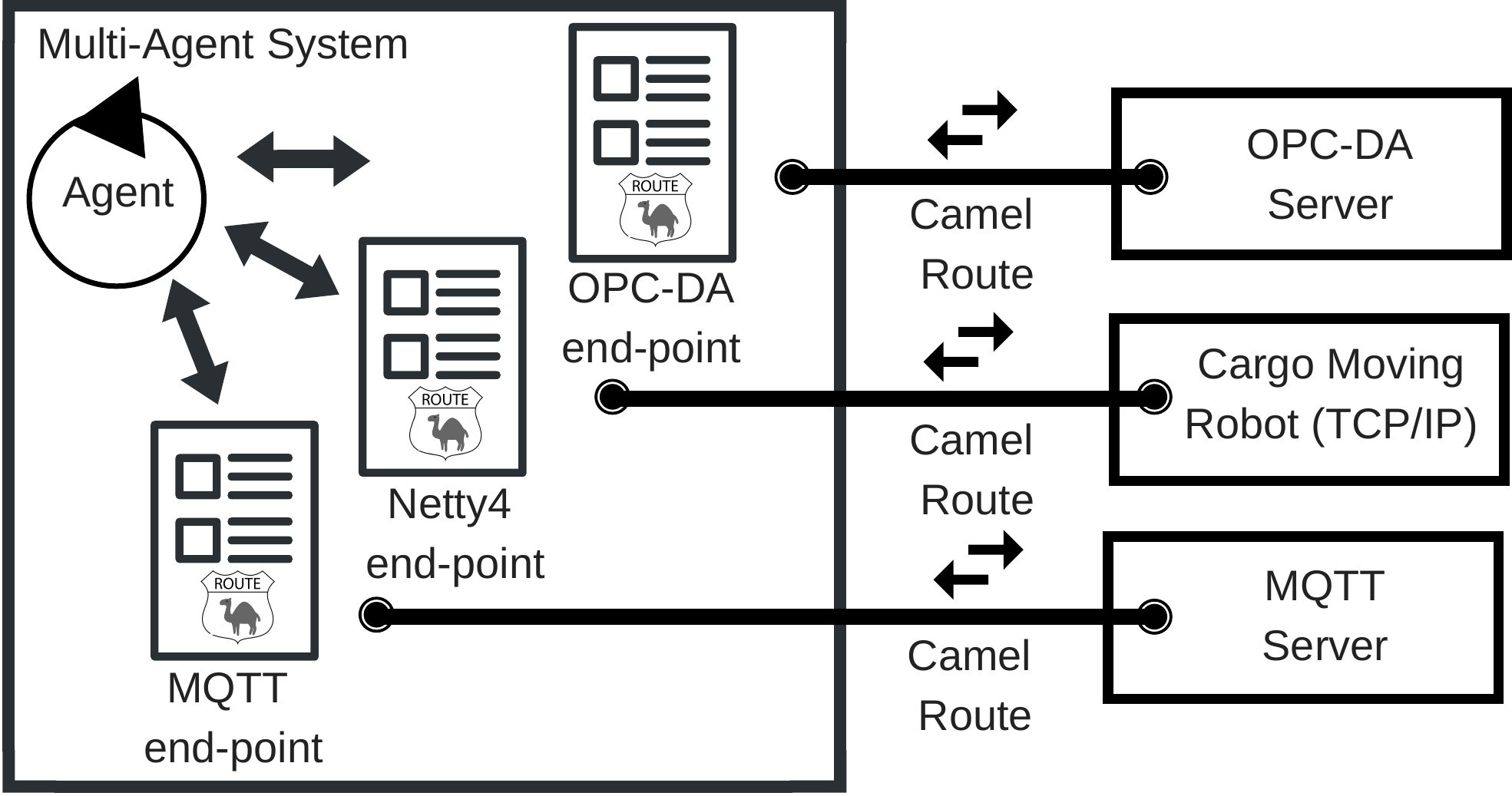}
}
\caption{MAS communicating with different resources.}
\label{fig:IndustrialEssay}
\end{figure}

The \emph{CamelArtifact} was tried in an Industry 4.0 context scenario (Figure~\ref{fig:IndustrialEssay}). To assess the solution, the application uses three resources with different features, one being an OPC\footnote{OPC: integration standard in industrial automation.} server, which may reflect some current industrial resource (e.g. a Programmable Logical Controller - PLC). Another resource is a compatible MQTT client, which may be an IoT device (e.g. a sensor). Finally, the last resource is a generic TCP/IP entity, which may be a software communicating by socket (e.g. a Robot or an Enterprise Resource Planning - ERP). In this test, all the three resources, were modelled as \emph{CamelArtifact}. 

To check the OPC-DA route a numerical variable was created in the OPC server and an observable property in the artifact. The OPC-DA component used is provided by an independent developer. The routes did the synchronization of the value of the variable with the observable property. The MQTT client, using the supported component, had its routes to send and receive messages from a MQTT broker. Finally, a robot firmware was deployed in a virtual machine to simulate a cargo moving robot. A generic TCP/IP route, using Netty4 supported endpoint, was set to send and receive messages in a proprietary protocol.

The MAS was designed with two Jason agents, one of them responsible to make the artifacts and share with another agent, a counter, the relevant information that came from the OPC server. The interoperability was tried when the agent used the counter from OPC server to act over the MQTT sensor and TCP/IP robot according to the variance of this information. 

\section{Related work}\label{sec:Related}

Among related works we are considered only MAS based studies. \cite{Mzahm2013} coined the term Agent of Things (AoT) which is IoT with reasoning capabilities. Their approach suggests the \emph{agentification} of things to reach the required intelligence, benefits and costs. However, it does not get use of all advantages of a multi-dimensional MAS, e.g. virtual and shareable entities accessible to autonomous actors. \emph{Agentification} tendency can be seen in other attempts to use MAS in the industry~\citep{Marik:2005:RAA:1082473.1082812,Yokogawa2000MachineryCS}, as in the research of \cite{Leitao2018}. The drawbacks of \emph{agentification} are the increase of complexity and reduction of scalability.

\cite{Maturana1996} have proposed a mediation and coordination tool for MAS. They have used mediator agents as manufacturing coordinators. Our proposal does not aims to put in the middle an autonomous entity; but it gives connectivity power in an A\&A MAS using a mainstream technology such as Camel. Following similar idea, \cite{Olaru2013} have developed an agent-based \emph{middleware} which creates a sub-layer of application layer that allows agents to mediate communications. Later other research tried to address mainly the coordination and organisation challenges~\citep{BARBOSA201599,KOTAK200395} regarding manufacturing scenarios. It is important to notice that the mediation functions of these works are limited comparing to Camel features. These approaches also lack an environment support as \textsf{CArtAgO} provides. 

\cite{TICHY2012846} used the Agent Development Environment (ADE) designed by Rockwell Automation, Inc. Besides the connectivity with the common shop floor devices (e.g. PLCs), this framework also supports the development of agents. They presented a conception to allow low and high level interaction, this last made by agents. The approach is an important industry supplier effort towards the requirements of the \emph{Smart Factory}. It also partly uses well-matured technology which is crucial for industrial stakeholders~\citep{Leitao2018}. The limitation we have seen regards specially connectivity with all sorts of entities (e.g. IoT sensors and mobile devices, ERP and other software, etc). Alternatively, the use of mature technologies can be reached using proper camel components to connect to industrial devices (e.g. using camel OPC-DA component).

\cite{Cranefield2013} developed a Camel component for \emph{Jason} agents. In this case environment description are not under A\&A concepts but they are part of the agents' knowledge. In their work, the agents are empowered by all Camel features we gave for \emph{CamelArtifacts}. Integration made only by agent dimension brings two advantages: (i) the programmer does not need to learn about artifacts, (ii) some elements are better modelled as agents, such as agents of another MAS. In contrast, two drawbacks: (i) agents increase computational demand higher than artifacts doing the same task, (ii) bring synchronization challenges, since agents spend time sharing information about the environment.

As far as we know, the only study that made use of A\&A concept to build MAS for industrial application in Industry 4.0 context was proposed by \cite{RoloffSHSPH14}. The limitation of this study refers to the integration by a unique API for OPC communication. The integration with other technologies needs the use of other APIs that brings more programming efforts when compared with our solution using Camel. Our proposal is filling a gap between the mature framework Apache Camel, a comprehensive mediation tool, and artifacts, a first-class designing entity.

\section{Conclusion}




In this paper, we discussed the interoperability challenge of the \emph{smart factory}. We showed that Agents and Artifacts (A\&A) method is useful for modelling the factory since this approach simplifies the design of non autonomous entities and may give more scalability to the whole system. We presented Camel framework as a mediation tool for integration with dozens of technologies used on industry. Our component, besides Camel facilities, also has functions to allow different topologies to deal with message rate requirements as well as resource limitations.


As future work we intend to change from polling strategy to event-driven strategy for consumer and producer sides. We intend to work on camel-agent~\citep{Cranefield2015} component trying to make both camel-artifact and camel-agent components with similar and easy to use configuration interfaces. Finally, we think that Apache Camel with both components may be used as Jason infrastructure being the mediation tool among distributed agents and artifacts.


\bibliography{sbaconf}

\end{document}